\documentclass[12pt]{article}
\usepackage{epsfig}
\usepackage[dvips]{color}
\begin{document}

\def\be{\begin{equation}}
\def\ee{\end{equation}}
\def\bq{\begin{equation}}
\def\eq{\end{equation}}
\def\bqa{\begin{eqnarray}}
\def\eqa{\end{eqnarray}}
\def\roughly#1{\mathrel{\raise.3ex
\hbox{$#1$\kern-.75em\lower1ex\hbox{$\sim$}}}}
\def\lsim{\roughly<}
\def\gsim{\roughly>}
\def\llgm{\left\lgroup\matrix}
\def\rrgm{\right\rgroup}
\def\vectrl #1{\buildrel\leftrightarrow \over #1}
\def\partrl{\vectrl{\partial}}
\def\gslash#1{\slash\hspace*{-0.20cm}#1}

\begin{center}
{\LARGE\bf
The Energy-Dependent Black-Disk Fraction\\
in Proton-Proton Scattering} \\
\end{center}
 \vspace {0.5 cm}
\begin{center}
{\bf Dieter Schildknecht} \\[2.5mm]
Fakult\"{a}t f\"{u}r Physik, Universit\"{a}t Bielefeld \\[1.2mm] 
D-33501 Bielefeld, Germany\\[1.2mm] 
and \\[1.2mm] 
Max-Planck-Institut f\"ur Physik (Werner-Heisenberg-Institut),\\[1.2mm] 
F\"ohringer Ring 6, D-80805 M\"unchen, Germany
\end{center}

\vspace{2 cm}

\begin{abstract}
Recent work by Block et al  on the  energy-invariant edge  in  proton-proton  
scattering   is  interpreted by quantitatively introducing a energy-dependent
black-disk fraction of the proton-proton
interaction.
\end{abstract}

In the context of the investigations on the empirical evidence for the
approach of the proton-proton-scattering cross section to a black-disk
limit at very high energies, the variable
\be
t = \frac{\sigma^{TOT} (W^2) - 2 \sigma^{EL} (W^2)}{\sqrt{\frac{\pi
\sigma^{TOT} (W^2)}{2}}}
\label{1}
\ee
was introduced \cite{Block1}, compare also refs. \cite{Block2,Block3,Stodolsky}.

The numerical values of $t$ can be determined from the experimental
results for the total $pp$ cross section, $\sigma^{TOT} (W^2)$, and
the elastic one, $\sigma^{EL} (W^2)$, measured at the center-of-mass
energy $W$. In an eikonal description of $pp$ scattering, $t$ stands for
the width of the edge of the scattering region in impact-parameter space
\cite{Block1}. Employing $t$, in the present note, we wish to
{\it explicitly} elaborate on how the variable $t$ specifies the fraction of
the total cross section due to black-disk scattering as a function of the
energy $W$.

From the experimental results on $\sigma^{TOT}$ and $\sigma^{EL}$, technically
by evaluating precise fits \cite{Block1}, \cite{Block2}, \cite{Block3},
\cite{Block4} to the experimental data for $\sigma^{TOT}$, $\sigma^{EL}$
and the inelastic cross section, $\sigma^{INEL}~(W^2)$, it was found that
$t$ in (\ref{1}) is independent of the energy $W$,
\be
t = const \simeq 1.1 fm.
\label{2}
\ee
From the extrapolation of the fits to energies far beyond the presently
accessible ones, the remarkable constancy of $t \simeq 1.1 fm$ in (\ref{2})
was found to remain valid in the huge range of $W \simeq 10 GeV$ to
$W \simeq 10^{13} GeV$ \cite{Block1}

The significance of the empirical observation (\ref{2}) is best
illuminated by rewriting (\ref{1}) as
\be
\sigma^{EL} (W^2) = \frac{1}{2} \sigma^{TOT} (W^2) \left( 1 -
\frac{t}{\sqrt{\frac{2 \sigma^{TOT} (W^2)}{\pi}}} \right).
\label{3}
\ee
Relation (\ref{3}) is seen to correspond to the transition from two
W-dependent observables $\left( \sigma^{TOT} (W^2), \sigma^{EL} (W^2)
\right)$ to the two observables $\left( \sigma^{TOT} (W^2), t \simeq
const \cong 1.1 fm \right).$
A measured value of the single quantity $\sigma^{TOT} (W^2)$ at the
energy $W$, together with the constant $t \cong 1.1 fm$, determines the
elastic cross section in (\ref{3}), as well as the inelastic one,
\be
\sigma^{INEL} (W^2) = \frac{1}{2} \sigma^{TOT} (W^2) \left( 1 +
\frac{t}{\sqrt{\frac{2 \sigma^{TOT} (W^2)}{\pi}}} \right).
\label{4}
\ee
Due to the slow growth of $\sigma^{TOT} (W^2) \simeq c_0 + c_2 \ln^2
(W^2/2 m^2_p)$, where $c_0 \simeq 24 mb = 2.4 fm^2$ and
$c_2 \simeq 0.23 mb = 0.023 fm^2$ 
 \cite{Block3}, for $W \to \infty$,
the black-disk limit
\be
\lim_{W \to \infty} \frac{\sigma^{EL} (W^2)}{\sigma^{TOT} (W^2)}
= \frac{1}{2}
\label{5}
\ee
in (\ref{3}) is reached only at energies far beyond the ones available in
the laboratory. Even for the very high energies explored at the Large
Hadron Collider (LHC) and by the Auger \cite{Abreu} and HiRes collaborations
\cite{Abbasi}, one finds \cite{Block3} values of $\sigma^{EL}/\sigma^{TOT}$
of 0.28 (at 14 TeV) and 0.30 (at 57 TeV) still significantly below
$\sigma^{EL}/\sigma^{TOT} = 0.50$ in (\ref{5}).

At finite energy $W$, the pair of observables $(\sigma^{TOT}(W^2),~t \simeq
{\rm const} \simeq 1.1 {\rm fm})$ according to (\ref{1}) determines the normalized
deviation of the total cross section from the black-disk limit (\ref{5}), 
\begin{eqnarray}
\frac{\sigma^{TOT} (W^2)- 2\sigma^{EL} (W^2)}{\sigma^{TOT}(W^2)}=
\frac{t}{\sqrt{\frac{2\sigma^{TOT} (W^2)}{\pi}}}   ,
\label{6}
\end{eqnarray}
as well as the normalized difference between the inelastic and the elastic cross
section given by
\begin{equation}
\frac{\sigma^{INEL} (W^2)-\sigma^{EL} (W^2)}{\sigma^{TOT} (W^2)} =
\frac{t}{\sqrt{ \frac{2\sigma^{TOT} (W^2)}{\pi}}}
\label{7}
\end{equation}
where $\sigma^{TOT} (W^2) = \sigma^{EL} (W^2) + \sigma^{INEL} (W^2)$ is used in
the transition from (\ref{6}) to (\ref{7}). The limit of $W\rightarrow\infty$
in (\ref{6}) takes
us back to (\ref{5}).

It is useful to {\it define} a black-disk total cross section, $\sigma^{TOT}_{\rm
  Black} (W^2)$, associated with a given value of the elastic cross section via 
\begin{equation}
\sigma^{TOT}_{\rm Black} (W^2) = 2 \sigma^{EL} (W^2).
\label{8}
\end{equation}
Inserting (\ref{8}) into (\ref{6}) yields 
\begin{equation}
\sigma^{TOT} (W^2) = \sigma^{TOT}_{\rm Black} (W^2) + t \sqrt{ \frac{\pi}{2}
  \sigma^{TOT} (W^2)}.
  \label{9}
\end{equation}
The total cross section is accordingly recognized as the sum of a contribution
that is associated with a black disk according to (\ref{8}),  
and a contribution that is proportional to the
constant $t$. The fraction of the total cross section originating from the
black-disk interaction according to (\ref{9}) is given 
\begin{equation}
\frac{\sigma^{TOT}_{\rm Black} (W^2)}{\sigma^{TOT} (W^2)} = 1 - \frac{t}{\sqrt{
\frac{2\sigma^{TOT} (W^2)}{\pi} }} = 
\left\{ \matrix{ & 1 ~ {\rm for}
\sqrt{\frac{2\sigma^{TOT} (W^2)}{\pi}} \gg t, \cr
& \frac{1}{2} ~ {\rm for} \sqrt{ \frac{2\sigma^{TOT} (W^2)}{\pi}} = 2t \cr} \right. .
\label{10}
\end{equation}
In (\ref{10}), besides the asymptotic value of unity (\ref{5}), we have given
the numerical
value of the ratio
$\sigma^{TOT}_{\rm Black} (W^2) / \sigma^{TOT} (W^2) = 1/2$ that is obtained
for $\sigma^{TOT} = 2 \pi t^2$. With the empirical value of $t = 1.1 fm$, this
value of $\sigma^{TOT} (W^2) = 2 \pi t^2$ coincides with the measurements of
$\sigma^{TOT} (W^2)$ at the energy of approximately $W \cong 2 TeV$ [1,2,3].
From (\ref{10}), at $W \cong 2 TeV$, the black disk defined by (\ref{8})
contributes about
one half of the sum on the right-hand side in (\ref{9}).
Or, equivalently, the cross section of a black disk determined by a given value
of $\sigma^{EL}(W^2)$ according to (\ref{8}), for $W=2$TeV is only half
as large as the
empirical cross section according to (\ref{9}).  

Additional insight into the significance of the representation (\ref{9}) is obtained
by introducing radii for the cross sections $\sigma^{TOT} (W^2)$ and
$\sigma^{TOT}_{\rm Black} (W^2)$, respectively, 
\begin{equation}
\sigma^{TOT} (W^2) = 2 \pi R_{TOT}^2 (W^2) ,
\label{11}
\end{equation}
as well as 
\begin{equation}
\sigma^{TOT}_{\rm Black} (W^2) = 2 \sigma^{EL} (W^2) = 2 \pi R^2_{\rm Black}
(W^2) ,
\label{12}
\end{equation}
In distinction from the black-disk radius in (\ref{12}), $R_{\rm Black}(W^2)$, the
radius $R_{TOT} (W^2)$ introduced in (\ref{11}) must be considered as an effective
radius. It approaches the radius of a black disk, $R_{TOT} (W^2) \rightarrow
R_{\rm Black} (W^2)$, in the asymptotic limit of $W\rightarrow \infty$, where
$\sigma^{TOT} (W^2) \rightarrow \sigma^{TOT}_{\rm Black} (W^2)$, compare (\ref{5})
and (\ref{8}). 

With (\ref{11}), the cross section (\ref{9}) is represented by
\begin{equation}
\sigma^{TOT} (W^2) = \sigma^{TOT}_{\rm Black} (W^2) + \pi t R_{TOT} (W^2).
\label{13}
\end{equation}
Note that (\ref{13}) is an empirical relation that only rests on the defining
equations (\ref{1}), (\ref{11}) and (\ref{12}) in terms of the measurable
cross sections
$\sigma^{TOT}(W^2) = \sigma^{EL} (W^2) + \sigma^{INEL} (W^2)$ and $\sigma^{EL}
(W^2)$. Note also that 
$\pi t R_{TOT} (W^2) = \pi t \sqrt{\sigma^{TOT}
  (W^2)/2\pi}$ is a measurable quantity.
The additive contribution $\pi t R_{TOT} (W^2)$ in (\ref{13}), properly called the
''edge'' \cite{Block1}, is recognized as a contribution
to $\sigma^{TOT} (W^2)$ that is
proportional to the area $2\pi t R_{TOT} (W^2)$ of a ring of constant thickness
(or width) $t = const$ centered around the radius $R_{TOT} (W^2)$. 
Multiplication by a factor $2 * 1/4$, where the factor 2 takes blackness into
account, and 1/4 is due to decreasing absorption \cite{Block1,Block2},
leads to $\pi t R_{TOT}
(W^2)$ in (\ref{13}). 
Returning to $\sigma^{INEL} (W^2)$ and $\sigma^{EL} (W^2)$ in
(\ref{13}), we see that the edge contribution yields the excess of $\sigma^{INEL}
  (W^2)$ over $\sigma^{EL}(W^2)$, 
\begin{equation}
\sigma^{INEL} (W^2) - \sigma^{EL} (W^2) = \pi t R_{TOT} (W^2) .
\label{14}
\end{equation}

The ratio (\ref{10}) of $\sigma^{TOT}_{\rm Black} (W^2)$ to $\sigma^{TOT} (W^2)$,  
with (\ref{11}) to (\ref{13}), at finite energy $W$
deviates from the asymptotic value of
unity by the normalized edge contribution. Explicitly, 
\begin{equation}
\frac{\sigma^{TOT}_{\rm Black} (W^2)}{\sigma^{TOT} (W^2)} = 1 - \frac{t}{2
  R_{TOT} (W^2)},
  \label{15}
\end{equation}
as well as 
\begin{equation}
\frac{R_{\rm Black} (W^2)}{R_{TOT} (W^2)} = \sqrt{1 - \frac{t}{2 R_{TOT}
    (W^2)}} = \left\{ \matrix{ 1 & {\rm for} R_{TOT}(W^2) \gg t , \cr
 \frac{1}{\sqrt 2} & \cong 0.71 {\rm for} R_{TOT}(W^2) = t } \right. .
 \label{16}
\end{equation}
At energies of approximately $W \cong 2$TeV, a region defined by an
interaction radius of about 71\% of the full interaction radius is associated
with a black-disk interaction, the remaining part is due to the
edge. Explicitly, 
\begin{equation}
\frac{\sigma^{TOT} (W^2) - \sigma^{TOT}_{\rm Black} (W^2)}{\sigma^{TOT} (W^2)}
= \frac{t}{2 R_{TOT} (W^2)} = \left\{ \matrix{ 0 , &{\rm for}~ R_{TOT}(W^2) \gg t ,
    \cr
\frac{1}{2} , & {\rm for}~ R_{TOT} (W^2) = t , } \right.
\label{17}
\end{equation} 
as well as 
\begin{equation}
\frac{R_{TOT} (W^2) - R_{\rm Black} (W^2)} {R_{TOT} (W^2)} = 1 - \sqrt{ 1 -
  \frac{t}{2 R_{TOT} (W^2)}} = \left\{ \matrix{ 
0 & {\rm for}~ R_{TOT}(W^2) \gg t \cr 
0.29 & {\rm for}~ R_{TOT}(W^2) = t } \right.
\label{18}
\end{equation}

\medskip\noindent
For $W \rightarrow \infty$, the relative contribution from the edge to (\ref{9}),
according to (\ref{17}) and (\ref{18}) tends to zero, the asymptotic
limit (\ref{3}) is reached. 

To gain further insight into the interplay betweeen the black-disk part of the
total cross section and the edge in (\ref{9}) and (\ref{13}), we go back to the
impact-parameter representation of the eikonal picture which actually was the
starting point when quantifying the concept of a smooth edge \cite{Block1}.
In impact-parameter space, denoting the transparency function by
$\eta \equiv \eta(b, R_{TOT}
(W^2))$, the representation (\ref{9}) of the total cross section reads
\cite{Block1, Block2}
\be
\sigma^{TOT} (W^2) = 4 \pi \int db~b [(1-\eta)^2 +
\eta (1-\eta)],
\label{19}
\ee
where, for simplicity of the present presentation, a real part of the scattering
amplitude is suppressed. The first and second parts on the right-hand side in
(\ref{19}) correspond to the first and second parts in (\ref{9}) and (\ref{13}).

The contribution to the total cross section that is due to the edge of
constant width $t$ at half maximum is approximately bounded by values of the
impact paramater $b$ in the range of 
\begin{equation}
R_{TOT} (W^2) - t \le b \le R_{TOT} (W^2) + t .
\label{20}
\end{equation}
The precise boundary depends on the exact shape of the transparency function.
For impact parameters $b$ outside the range (\ref{20}), in particular for
values of $b$ approximately given by 
\begin{equation}
b \le R_{TOT} (W^2) - t \equiv b_{\rm Max} (W^2),
\label{21}
\end{equation} 
the contribution of the edge to the total cross section in (\ref{9}), (\ref{13})
and (\ref{19}) is
vanishing. The total cross section obtained by integration over $b$ up to $b
\le b_{\rm Max} (W^2)$ is approximately given by the black-disk value of 


\begin{equation}
\sigma^{TOT} (W^2)|_{b_{\rm Max} (W^2)} = 2 \sigma^{EL} |_{b_{\rm Max} (W^2)} = 2
\pi (R_{TOT} (W^2) - t)^2 .
\label{22}
\end{equation}
The fraction of the total cross section supplied by $\sigma^{TOT} (W^2)
|_{b_{\rm Max} (W^2)}$ becomes 
\begin{eqnarray}
& & \frac{\sigma_{TOT} (W^2) |_{b_{\rm Max} (W^2)}}{\sigma^{TOT} (W^2)} =  
\left( 1 - \frac{t}{R_{TOT}(W^2)} \right)^2 =  \nonumber\\ 
& & \nonumber \\
& & \left\{ \matrix{ 
1 & {\rm for}~ R_{TOT} (W^2) \gg t , \cr
0.063 & {\rm for}~ R_{TOT} (W^2) = R_{TOT} ( W^2 = (57 {\rm TeV})^2) = 1.47 fm, \cr
0.028 & {\rm for}~ R_{TOT} (W^2) = R_{TOT} ( W^2 = (14 {\rm TeV})^2) = 1.32 fm
. }
\right.
\label{23}
\end{eqnarray}
Even at the highest presently accessible energies, the range in
impact-parameter space of $b \le
b_{\rm Max} (W^2)$, where $\sigma^{TOT} (W^2) |_{ b_{\rm Max} (W^2)} = 2
\sigma^{EL} |_{b_{\rm Max} (W^2)}$,
according to (\ref{21}) and  (\ref{23}) is restricted to very small values
of $b_{\rm Max} \cong 0.2 fm$ to $0.4 fm$. Compare also refs. \cite{Block2} and
\cite{Alkin}.

\begin{figure}
\begin{center}
\epsfig{file=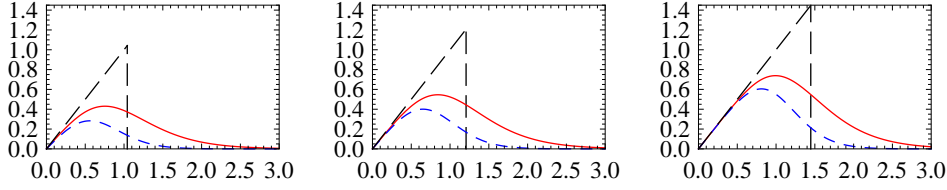}
\end{center}
\caption{The figures show amplitudes as a function of the impact parameter
$b$ for energies of $W = 1$ TeV, 5 TeV and 50 TeV. Integration over $b$ and
multiplication by $4\pi$ yields the total cross section
$\sigma^{TOT}(W^2)$(solid red curve), the cross section $\sigma^{TOT}_{\rm
  Black}(W^2) = 2 \sigma^{EL} (W^2)$ (short dashed blue curve) and the cross
section of a black disk of radius $R_{TOT} (W^2)= \sqrt{\sigma^{TOT}(W^2) / 2
  \pi}$ (long-dashed black curve). The vertical and horizontal scales give the
amplitude and $b$ in $fm$. The integral over the amplitude corresponding to
$\sigma^{TOT}(W^2)$ is equal to the integral over the black-disk amplitude
associated with radius $R_{TOT} (W^2)$. The region between the
$\sigma^{TOT}(W^2)$ and $\sigma^{TOT}_{\rm Black}(W^2)$ amplitudes is due to the
edge. This figure is based on the eikonal fits in reference \cite{Block2}, and it was
prepared by Phuoc Ha for the author of the present work.}
\end{figure}

Figure 1\footnote{The author thanks Phuoc Ha for preparing and 
 providing Fig. 1 based on the eikonal  fits from reference \cite{Block2}.}
shows eikonal fits  of scattering amplitudes in impact-parameter space for
  various energies $W$ including the extrapolation to $W = 50 TeV$. 
The integrals over the impact parameter $b$ of the amplitudes shown in Fig.1,
  upon multiplication by $4\pi$, yield fits to the experimental data for
  $\sigma^{TOT} (W^2)$ and $\sigma^{TOT}_{\rm Black} (W^2) = 2\sigma^{EL} (W^2)$. 
The Figure also shows a linearly rising theoretical black-disk amplitude. Upon
  integration, restricted by a chosen upper limit on $b$ and multiplication by
  $4\pi$, one obtains the corresponding cross section of a black disk. 
The Figure nicely displays the deviation between
  $\sigma^{TOT} (W^2) = 2 \sigma^{EL} (W^2) =
  \sigma^{TOT}_{\rm Black} (W^2)$ and
  $\sigma^{TOT} (W^2) = \sigma^{TOT}_{\rm Black} (W^2) + \pi t R_{TOT} (W^2)$ for
  impact parameters $b \ge b_{\rm Max} (W^2)$ due to contributions from the
  edge. For $b > R_{TOT} (W^2)$, the total cross section is dominantly due to the
  edge, and the edge is dominated by inelastic scattering.
  For $R_{TOT} (W^2) \rightarrow \infty$, the relative contribution of the edge
to the total cross section
goes to zero, and $\sigma^{TOT}_{\rm Black} (W^2) / \sigma^{TOT}
 (W^2)\rightarrow 1$.
 The huge energy required to approach this saturation limit of full blackness
 is inherently connected with both, the fairly large value of the edge
 constant $t \simeq 1.1 fm$, and the slow increase of $R_{TOT} (W^2)$ with
 the energy $W$ that is due to the small coefficient $c_2$ of the $\ln^2 (W^2)$
 term in the representation of the total cross section, $\sigma^{TOT} (W^2)$.
 We note that a (theoretical) value of $t \to 0$, corresponding
 to a black-disk interaction at all energies, is not only incompatible with the
 experimental value of $t \simeq 1.1 fm$, but it is also inconsistent with the
 original prediction and theoretical explanation of the $\ln^2 (W^2)$ dependence
 of the total cross section which relies \cite{Heisenberg} on an
 increase of the active
 interaction region, when increasing the energy.

The edge constant $t \simeq 1.1 fm$ determines the partition of the total
cross section $\sigma^{TOT} (W^2)$ for given $W$ into an elastic and an inelastic
part, and it quantifies the approach to the black-disk limit
\be
\lim_{W \to \infty} \frac{R_{Black} (W^2)}{R_{TOT} (W^2)} = 1
\label{24}
\ee
as a function of the $pp$ energy $W$. It seems appropriate to explcitly
elaborate on these interpretations of the work in (\ref{1}).

\bigskip

\leftline{\large \bf Acknowledgement}

This note was inspired by a lecture given  by Leo Stodolsky at the
Max Planck Institut f{\"u}r Physik in M{\"u}nchen.
The author thanks Leo Stodolsky for useful discussions and Phuoc Ha
for providing Figure 1.


\begin{thebibliography}{99}
\bibitem{Block1}
M.M. Block, L. Durand, F. Halzen, L. Stodolsky and T.J. Weiler,
``Evidence for an energy-invariant `edge' in proton-proton scattering
at very high energies'', Phys. Rev. D {\bf 91}, 011501 (R) (2015).

\bibitem{Block2}
M.M. Block, L. Durand, P. Ha, F. Halzen,
``Eikonal fit to $pp$ and $\bar p p$ scattering and the edge in the
scattering amplitude'', Phys. Rev. D {\bf 92}, 014030 (2015).

\bibitem{Block3}
M.M. Block, L. Durand, P. Ha, F. Halzen,
``Comprehensive fits to high energy data for $\sigma, \gamma$ and $B$ and
the asymptotic black-disk limit'', Phys. Rev. D {\bf 92}, 114021 (2015).

\bibitem{Stodolsky}
L. Stodolsky,
``Behavior of very High Energy hadronic cross sections'',
arXiv: 1703.05668 [hep-ph], Mod. Phys. Lett. A32 (2017) no. 31,  1730028.

\bibitem{Block4}
M.M. Block and F. Halzen,
``Analyticity as a robust constraint on the total cross section at the
CERN Large Hadron Collider'', Phys. Rev. D {\bf 73}, 054022 (2006).

\bibitem{Abreu}
P. Abreu et al. (Pierre Auger Collaboration), Phys. Rev. Lett.
{\bf 109}, 062002 (2012).

\bibitem{Abbasi}
R. Abbasi et al. (HiRes Collaboration), Astrophys. J. {\bf 684} (2008)
790.

\bibitem{Alkin}
A. Alkin, E. Martynov, O. Kovalenko and S.M. Troshin, Phys. Rev. D
{\bf 89}, 0991501 (2014).

\bibitem{Heisenberg}
W. Heisenberg, Vortr\"age \"uber kosmische Strahlung (Springer, Berlin 1953),
p. 155, reprinted in W. Heisenberg, Collected Works, Series B
(Springer, Berlin, 1984) p. 498;\hfill\break
W. Heisenberg, Die Naturwissenschaften 61 (1974), 1, reprinted in Collected
Works, Series B, p. 912.

\end{thebibliography}
\end{document}